\documentclass{PoS}

\usepackage{amsmath,amssymb}
\usepackage{wrapfig}

\title{The top-quark pair production cross section at next-to-next-to-leading logarithmic order}

\ShortTitle{The top-pair cross section at NNLL order}

\author{Martin Beneke\\
        Institute f\"ur Theoretische Teilchenphysik und 
Kosmologie,\\
RWTH Aachen University, D--52056 Aachen, Germany
}

\author{\speaker{Pietro Falgari} \footnote{Preprint numbers: TTK-11-60,
ITP-UU-11/45, SPIN-11/35, FR-PHENO-2011-024, 
SFB/CPP-11-79}\\
        Institute for Theoretical Physics and Spinoza Institute,\\
Utrecht University, 3508 TD Utrecht, The Netherlands\\
        E-mail: \email{p.falgari@uu.nl}}

\author{Sebastian Klein\\
        Institute f\"ur Theoretische Teilchenphysik und 
Kosmologie,\\
RWTH Aachen University, D--52056 Aachen, Germany
        }

\author{Christian Schwinn\\
       Albert-Ludwigs Universit\"at Freiburg, 
Physikalisches Institut, \\
D-79104 Freiburg, Germany
      }

\abstract{We present predictions for the total $t \bar{t}$ production 
cross section $\sigma_{t\bar t}$ at the Tevatron and LHC, which
include the resummation of soft logarithms and Coulomb singularities 
through next-to-next-to-leading logarithmic order, and $t \bar{t}$ 
bound-state contributions. Resummation effects amount to about
$8\%$ of the next-to-leading order result at Tevatron and about $3 \%$
at LHC with 7 TeV centre-of-mass energy. They
lead to a significant reduction of the theoretical uncertainty.   
With $m_t=173.3\,$GeV, we find 
\begin{equation}
\sigma_{t \bar{t}}^{\text{Tevatron}}=7.22^{+0.31+0.71}_{-0.47-0.55}\,\, 
\text{pb} 
\hspace{2 cm} 
\sigma_{t \bar{t}}^{\text{LHC}}=162.6^{+7.4+15.4}_{-7.5-14.7}\,\, 
\text{pb} \, ,
\nonumber
\end{equation}
in good agreement with the latest experimental measurements.}

\FullConference{ 10th International Symposium on Radiative Corrections (Applications of Quantum Field Theory to Phenomenology) - Radcor2011\\
September 26-30, 2011\\
Mamallapuram, India}

\begin{document}

\section{Introduction}

The total top-pair production cross section $\sigma_{t \bar{t}}$ has been measured at Tevatron with an accuracy 
$\Delta \sigma_{t \bar{t}}/\sigma_{t \bar{t}}$ of about $\pm 7\%$
\cite{arXiv:1004.3224, arXiv:1101.0124}
and the two LHC experiments have already reached similar sensitivity \cite{Aad:2011yb,Chatrchyan:2011yy}, 
with the accuracy of some analyses already at the $\pm 6.5\%$ level. With
more statistics being collected, the experimental error is bound to
reduce even further, 
opening a new era of precision top phenomenology. 
The total cross section, in particular, can be used to measure the top-quark pole mass $m_t$ in a theoretically clean way, 
to constrain new physics
and to extract information on the gluon distribution function (PDF) of the proton. 
This clearly requires a precise theoretical understanding of the $t \bar{t}$-production dynamics;
more specifically, predictions beyond next-to-leading order (NLO) in standard fixed-order perturbative QCD are necessary. 

Near the partonic production threshold, $\beta \sim 0$, where $\beta\equiv \sqrt{1-4 m_t^2/\hat{s}}$ is the
velocity of the two top quarks, the partonic cross sections $\hat{\sigma}_{pp' \rightarrow t \bar{t}X}$ are 
enhanced due to suppression of soft-gluon emission and exchange of
potential (Coulomb) gluons between the non-relativistic top and anti-top. These two effects give rise
to singular terms, at all orders in perturbation theory, of the form $\alpha_s \ln^{2,1} \beta$ and $\alpha_s/\beta$, respectively. 
While the hadronic cross section,
\begin{equation} \label{eq:hadron}
\sigma_{t \bar{t}}(s)=\sum_{p,p'=q,\bar{q},g} \int_{4 m_t^2/s}^1 d \tau L_{pp'}(\tau,\mu_f) \hat{\sigma}_{pp' \rightarrow t \bar{t}X} (\tau s,\mu_f) \, ,
\end{equation} 
where $L_{pp'}$ are parton luminosity functions, receives contributions from regions where $\beta$ is not necessarily small, especially for LHC centre-of-mass
energies, the region with $\beta \lesssim 0.3$ still gives a sizeable contribution to $\sigma_{t \bar{t}}$, due to the rapid
fall off of the parton luminosity functions at large $\tau$. One therefore argues that
an all-order resummation of soft and Coulomb corrections provides more accurate
predictions than the fixed-order calculation.

Leading logarithmic (LL) and next-to-leading logarithmic (NLL) resummation in the so-called Mellin-space formalism  
has been known for a while \cite{Laenen:1991af}. 
More recently, thanks also to the calculation of the
relevant soft anomalous dimensions \cite{arXiv:0907.1443}, next-to-next-to-leading logarithmic (NNLL) resummations
\cite{Ahrens:2011mw,arXiv:1109.1536,arXiv:1111.5869}, 
and approximated NNLO cross sections constructed from the re-expansion of the
resummed result \cite{arXiv:0911.5166,arXiv:1007.1327,Ahrens:2011px,arXiv:1109.3231}, have become available. 
Of the aforementioned works, only Ref.~\cite{arXiv:1109.1536} provides a simultaneous resummation of soft and Coulomb corrections, 
obtained through a general formalism  \cite{arXiv:1007.5414} based on the factorization of soft and Coulomb effects
in the context of soft-collinear effective theory (SCET) and potential non-relativistic QCD (PNRQCD).
The formalism, and the results of~\cite{arXiv:1109.1536}, will be reviewed in the following. 
 
\section{Factorization and resummation of the $t \bar{t}$ total cross section}

The basis for resummation is the factorization of the partonic cross section into short-distance contributions, related to
physics at the hard scale $\sim m_t$, and effects associated with soft-gluon emission, which naturally "live" at a 
much smaller scale $\sim m_t \beta^2$. In \cite{arXiv:1007.5414} it was shown that, at threshold, 
the partonic cross section in fact factorizes into three different contributions,    
\begin{equation}
\label{eq:fact}
  \hat\sigma_{pp'\rightarrow t \bar{t}X }(\hat s,\mu)
= \sum_{R={\bf 1},{\bf 8}}H^{R}_{pp'}(m_t,\mu)
\;\int d \omega\;
J_{R}(E-\frac{\omega}{2})\,
W_i^{R}(\omega,\mu)\, .
\end{equation}
The hard function $H^{R}_{pp'}$ encodes the model-specific short-distance effects, the Coulomb 
function $J_{R}$ describes the internal evolution of the $t \bar{t}$ pair, driven by Coulomb exchange,
and the soft function $W_i^{R}$ contains soft-gluon contributions. Here $R$ denotes the irreducible 
colour representation, either singlet ($R={\bf 1}$) or octet ($R={\bf 8}$), of the $t \bar{t}$ pair.

In the approach adopted here \cite{arXiv:1007.5414,arXiv:0710.0680}, the hard and soft functions, 
$H^{R}_{pp'}$ and $W_i^{R}$, are 
resummed through renormalization-group (RG) evolution equations directly in momentum space, 
contrary to the conventional formalism, where resummation is performed in Mellin-moment space. 
The relevant RG equations and their solutions were derived, for the
general case of massive particle pairs $H H'$ in arbitrary colour representations $R,\,R'$ , in \cite{arXiv:1007.5414}. 
For the soft function $W_i^{R}$ the solution to the evolution equation reads
\begin{equation}
\label{eq:w-resummed}
W^{R,\text{res}}_{i}(\omega,\mu)=
\exp[-4 S(\mu_s,\mu)+2 a^{R}_{W,i}(\mu_s,\mu)]\,
\tilde{s}_{i}^{R}(\partial_\eta,\mu_s) 
\frac{1}{\omega} \left(\frac{\omega}{\mu_s}\right)^{2 \eta} \theta(\omega)
\frac{e^{-2 \gamma_E \eta}}{\Gamma(2 \eta)}\, .
\end{equation}
In (\ref{eq:w-resummed}), $\tilde{s}_{i}^{R}$ represents the Laplace transform of the fixed-order soft function $W_i^{R}$.
The function $S$ controls the resummation of double logarithms, while $a^{R}_{W,i}$ and $\eta$ resum single logarithms. 
To obtain NNLL accuracy, these functions must be included at the three-loop ($S$) or two-loop ($a^{R}_{W,i}$ and $\eta$)
order, while the fixed-order soft function $\tilde{s}_{i}^{R}$ is required at one loop. 
An expression analogous to (\ref{eq:w-resummed}) can be derived for the hard function $H_{pp'}^R$. 

Resummation of Coulomb effects has been extensively studied in the context of PNRQCD and quarkonia physics.
The potential function $J_{R}$ is related to the Green's function of the Schr\"odinger operator
$-\vec{\nabla}^{\,2}/m_t-(-D_{R})\,\alpha_s/r\, [1+{\cal O}(\alpha_s)]$,
\begin{equation}
J_{R}(E)=2 \,\mbox{Im} \left[\,
G^{(0)}_{C,R}(0,0;E) \,\Delta_{\rm nC}(E) + G^{(1)}_{C,R}(0,0;E) + 
\ldots\right] \, . 
\label{JRal}
\end{equation}
In Eq. (\ref{JRal}), $G^{(0)}_{C,R}$ denotes the LO Coulomb Green's function, 
\begin{equation}
G^{(0)}_{C,R}=-\frac{m_t ^2}{4 \pi} \Bigg\{\!
    \sqrt{-\frac{E}{m_t}}+ (-D_{R})\alpha_s \bigg[
      \frac{1}{2}\ln \bigg(\! -\!\frac{4 \,m_t E}{\mu_C^2}\!\bigg)\! -\frac{1}{2}  +\gamma_E
    +\psi\bigg(1-\frac{(-D_{R})\alpha_s}{
      2\sqrt{-E/ m_t}}\!\bigg)\!\bigg]\!\Bigg\} \, ,
\end{equation}
with $E=\sqrt{\hat{s}}-2 m_t \sim m_t \beta^2$. $G^{(1)}_{C,R}$ is the correction to the Green function from the $\alpha_s^2$ terms in the Coulomb potential,
and $\Delta_{\rm nC}=1+{\cal O}(\alpha_s^2 \ln \beta)$ represents non-Coulomb contributions which enter the cross section first at NNLL order \cite{arXiv:0911.5166}.
Notice that for $E<0$, i.e. below the production threshold, 
Eq. (\ref{JRal}) contains a series of $t \bar{t}$ bound-state
resonances when $D_R<0$, which are included in the NNLL results presented in Section \ref{sec:results}. 

It is often convenient to re-expand the resummed results to construct higher-order approximations at fixed order in $\alpha_s$. 
In particular, at NNLL all singular terms in the limit $\beta \rightarrow 0$ at ${O} (\alpha_s^4)$ can be correctly predicted. Hence, one can define an approximated NNLO
prediction as
\begin{equation}
\hat{\sigma}_{pp'}^{\text{NNLO}_\text{app}}=\sum_R \left\{\hat{\sigma}_{pp',R}^{\text{NLO}}+\sigma_{pp',R}^{(0)} \left(\frac{\alpha_s}{4 \pi}\right)^2 \sum_{n=0,2} f_{pp',R}^{(2,n)} \ln^n \frac{\mu_f}{m_t} \right\} \, ,
\label{eq:NNLO}
\end{equation}
with $\hat{\sigma}_{pp',R}^{\text{NLO}}$ the exact colour-separated
NLO cross sections \cite{Czakon:2008ii}, and $\sigma_{pp',R}^{(0)}$ the
Born contributions. The NNLO functions $f_{pp',R}^{(2,n)} $  
incorporate exactly all terms (and only those) of the form
$\ln^{4,3,2,1} \beta$, $\frac{\ln^{2,1} \beta}{\beta}$,
$\frac{1}{\beta^{2,1}}$, and were first correctly obtained in
\cite{arXiv:0911.5166}.

\section{Scale choices and theoretical uncertainties} 
\label{sec:scales}

The starting point of the evolution of the soft function $W^{R}_i$, represented by the soft scale $\mu_s$ appearing in (\ref{eq:w-resummed}), 
has to be chosen such that logarithms in the expansion in $\alpha_s$ of $\tilde{s}_{i}^{R}$ are small, giving a stable perturbative behaviour at the low scale $\mu_s$.
For soft interactions, the natural scale is set by the kinetic energy of the $t \bar{t}$ pair, $m_t \beta^2$.  
 Accordingly, in \cite{arXiv:1109.1536} the soft scale in the resummed cross section was set to 
\begin{equation}
\mu_s = \text{max} [k_s m_t \beta_\text{cut}^2, k_s m_t \beta^2] \,  ,
\end{equation}
with the constant $k_s$ chosen by default as $k_s=2$. In the upper interval, $\beta>\beta_\text{cut}$, the soft logarithms $\ln \beta$ in the \emph{partonic} cross section
are correctly resummed by the 
running soft scale $k_s m_t \beta^2$. 
If $\beta_\text{cut}$ is not too big, in the lower interval, $\beta<\beta_\text{cut}$, the frozen soft scale $k_s m_t \beta_\text{cut}^2$ still correctly resums
the dominant contributions ($\sim \ln \beta_\text{cut}$) to the \emph{hadronic} cross section, 
at the same time avoiding ambiguities related to the Landau pole 
in $\alpha_s$. A prescription for the choice of $\beta_\text{cut}$ is 
detailed in \cite{arXiv:1109.1536}.

Contrary to the soft function, the hard function $H_{pp'}^R$ naturally lives at the parametrically larger scale given by the invariant mass of the $t \bar{t}$ pair. The default value for the hard scale $\mu_h$ appearing in the
resummed hard function is therefore chosen as $\mu_h =2 m_t$. 
On the other hand, the scale of Coulomb interactions is set by the typical virtuality of potential gluons, $q^2 \sim m_t^2 \beta^2$. For the Coulomb scale $\mu_C$ in 
(\ref{JRal}) we thus choose 
\begin{equation}
\mu_C = \text{max}[C_F \alpha_s m_t,2 m_t \beta] \, , 
\end{equation}
where the frozen scale at low values of $\beta$, equal to the inverse
Bohr radius $C_F \alpha_s m_t$, signals the onset of $t \bar{t}$ bound-state effects. 

There is clearly some degree of ambiguity in the choice of the scales appearing in the resummed result. Besides these, NNLL and approximated NNLO predictions
are affected by uncertainties related to constant and power suppressed terms which are not controlled by resummation. Therefore, to reliably ascertain the residual
theoretical error of the predictions presented below, we consider the following sources of ambiguity:
\begin{description}
\item[Scale uncertainty:] we vary all scales $\mu_i$  in the interval
  $[\tilde\mu_i/2,2\tilde \mu_i]$ around their central values
  $\tilde\mu_i$. $\mu_C$ is varied while keeping
  the other scales fixed. $\mu_h$ and $\mu_f$ are allowed to
  vary simultaneously,
  imposing the additional constraint $1\le \mu_h/\mu_f \le 4$.
  For the fixed-order results (NLO and NNLO$_\text{app}$) the factorization scale $\mu_f$
  and renormalization scale $\mu_r$ are varied simultaneously, with the constraint  
 $1/2\le \mu_r/\mu_f \le 2$.
  The errors from varying $\{\mu_f,\mu_h\}$ 
  and $\mu_C$ 
  are added in quadrature.   
\item[Resummation ambiguities:] we consider three different sources
  of ambiguities: i) the difference between the default setting 
  $E = m_t\beta^2$ compared to $E=\sqrt{\hat s}-2m_t$ in 
  (\ref{JRal}), ii) the difference between the NNLL
  implementation for the soft scale choices $k_s=1,4$
  to the default choice
  $k_s=2$, iii) the effect of varying 
  $\beta_{\text{cut}}$ by $20\%$ around the default
  value for $k_s=2$ (see  \cite{arXiv:1109.1536} for details). The resulting errors are
  added in quadrature.
  \item[NNLO-constant:] by default, the ${\cal O}(\alpha_s^4)$ constant in (\ref{eq:NNLO}), $C^{(2)}_{pp',R}$,  is set to zero. We estimate the effect
  of a non-vanishing constant by considering variations $C^{(2)}_{pp',R} = \pm |C^{(1)}_{pp',R}|^2$, with  
  $C^{(1)}_{pp',R}$ the constants in the threshold expansion of the NLO cross sections.
\end{description}
In Section \ref{sec:results} the total theory error is obtained summing in quadrature the three aforementioned uncertainties. 
Additionally, we estimate the error due to uncertainties in the PDFs and the strong coupling, 
using the $90\%$ confidence level set of the MSTW08NNLO 
PDFs and the five sets for variations of $\alpha_s$ 
provided in~\cite{Martin:2009iq}. This will be quoted separately from the theoretical error. 
   
\section{Results}
\label{sec:results}

We define our default prediction, NNLL$_2$, by matching the NNLL resummed result to the approximated NNLO cross section, i.e.
\begin{equation}
\text{NNLL}_2=\text{NNLL}-\text{NNLL}(\alpha_s^4)+\text{NNLO}_{\text{app}} \, ,
\end{equation} 
where $\text{NNLL}(\alpha_s^4)$ is the expansion, up to order $\alpha_s^4$, of the resummed result. The top-quark 
mass is set to $m_t=173.3 \text{GeV}$, and the renormalization and factorization scale are chosen as
$\mu_f=\mu_r=m_t$. The other scales are treated as explained in Section \ref{sec:scales}. For the convolution of the 
partonic cross section with the parton luminosity functions, Eq. (\ref{eq:hadron}), we use the MSTW08 PDF set~\cite{Martin:2009iq}.  

Numerical results for the cross section at the Tevatron and LHC for NLO, NNLO$_\text{app}$ and the resummed NNLL$_2$ implementation are given in Table \ref{tab:res}, with the 
first error corresponding to the total theoretical uncertainty, computed as described in Section \ref{sec:scales}, and the 
second error the PDF+$\alpha_s$ uncertainty. The genuine NNLL
corrections are sizeable and positive both at Tevatron ($\sim +13 \%$) and LHC ($\sim +9 \%$), 
though their effect is partially compensated by switching from NLO to NNLO PDF sets for the NNLO$_\text{app}$ and NNLL$_2$ results, which
results in a negative shift of the cross section. The bulk of the corrections beyond NLO is accounted for by the ${\cal O}(\alpha_s^4)$ terms, as evident from a
comparison of NNLO$_\text{app}$ and NNLL$_2$. One can also notice a significant reduction of the theoretical uncertainty from NLO to NNLO$_\text{app}$/NNLL$_2$, 
both at Tevatron
and the LHC, to the extent that for NNLO$_\text{app}$ and NNLL$_2$ the total error is dominated by the PDF+$\alpha_s$ error.

\begin{table}[t!]
\begin{center}
\begin{tabular}{|l|c|c|c|}
\hline
$\sigma_{t \bar{t}}$[pb]&  Tevatron
& LHC ($\sqrt{s}=$7 TeV) & LHC ($\sqrt{s}=$14 TeV) \\
\hline
\hline
NLO & $6.68^{+0.36+0.51}_{-0.75-0.45}$ & $158.1^{+18.5+13.9}_{-21.2-13.1}$ & $884^{+107+65}_{-106-58}$ \\
\hline
NNLO$_\text{app}$ & $7.06^{+0.27+0.69}_{-0.34-0.53}$ & $161.1^{+12.3+15.2}_{-11.9-14.5}$&  $891^{+76+64}_{-69-63}$ \\
\hline
NNLL$_2$ & $7.22^{+0.31+0.71}_{-0.47-0.55}$ & $162.6^{+7.4+15.4}_{-7.5-14.7}$&  $896^{+40+65}_{-37-64}$ \\
\hline
\end{tabular}
\end{center}
\caption{$t \bar{t}$ cross section at Tevatron and LHC from NLO,
  NNLO$_\text{app}$ and NNLL$_2$ approximations, for $m_t=173.3\,$GeV. The first error set denotes 
the total theoretical uncertainty, the second the PDF+$\alpha_s$ error.}
\label{tab:res} 
\end{table} 

\begin{figure}[t!]
\begin{center}
\includegraphics[width=0.48 \linewidth]{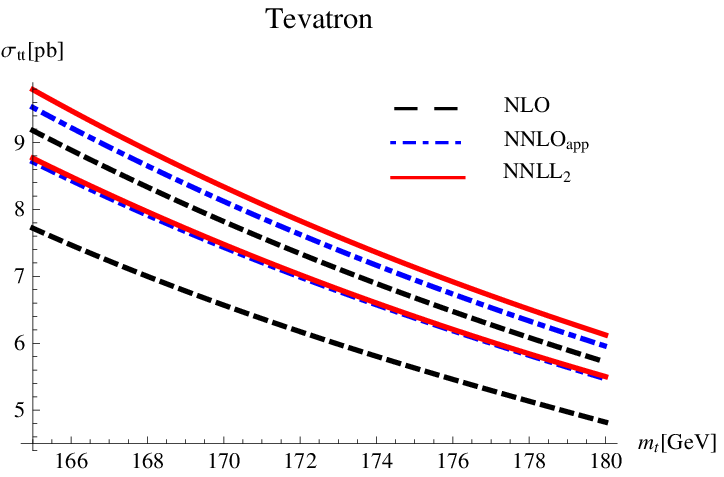}
\includegraphics[width=0.48 \linewidth]{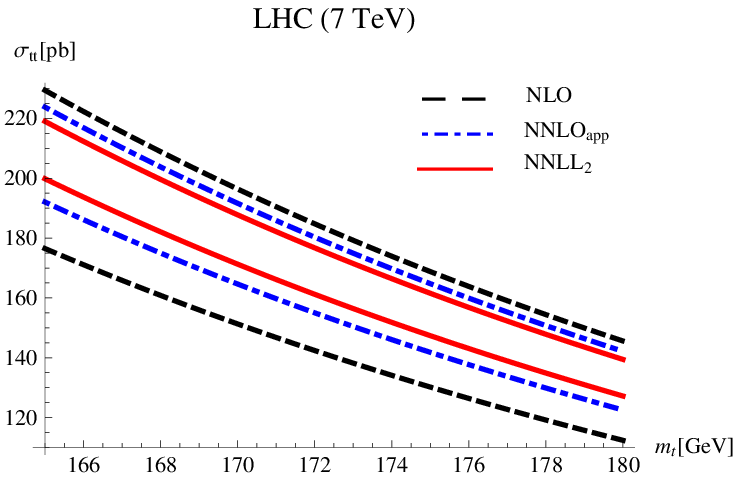}
\end{center}
\caption{NLO (dashed black), NNLO$_\text{app}$ (dot-dashed blue) and
  NNLL$_2$ (solid red) 
  as function of $m_t$. The bands
  correspond to the total theoretical uncertainty  (that is, excluding
  the PDF+$\alpha_s$
  error) of the prediction.}
\label{fig:theoryerror}
\end{figure}

The theory uncertainty of different approximations is shown in Figure \ref{fig:theoryerror}, where
the total theoretical error bands for NLO (dashed black), NNLO$_\text{app}$ (dot-dashed blue) and NNLL$_2$ (solid red) are plotted
as functions of $m_t$. At the LHC, one observes a nice convergence of the series NLO$\rightarrow$NNLO$_\text{app}$$\rightarrow$NNLL$_2$,
with the resummed result having the smallest theoretical uncertainty,
corresponding to about $\pm 4.7 \%$. At the Tevatron, however, the
NNLO$_\text{app}$ result exhibits the smallest error. We interpret this as 
an indication that the NNLO$_\text{app}$ error is accidentally 
underestimated by the scale variation procedure. For NNLL$_2$, the residual theoretical uncertainty is $+4.3\, -6.5 \%$.  
     
\begin{figure}[t!]
\begin{center}
\includegraphics[width=0.48 \linewidth]{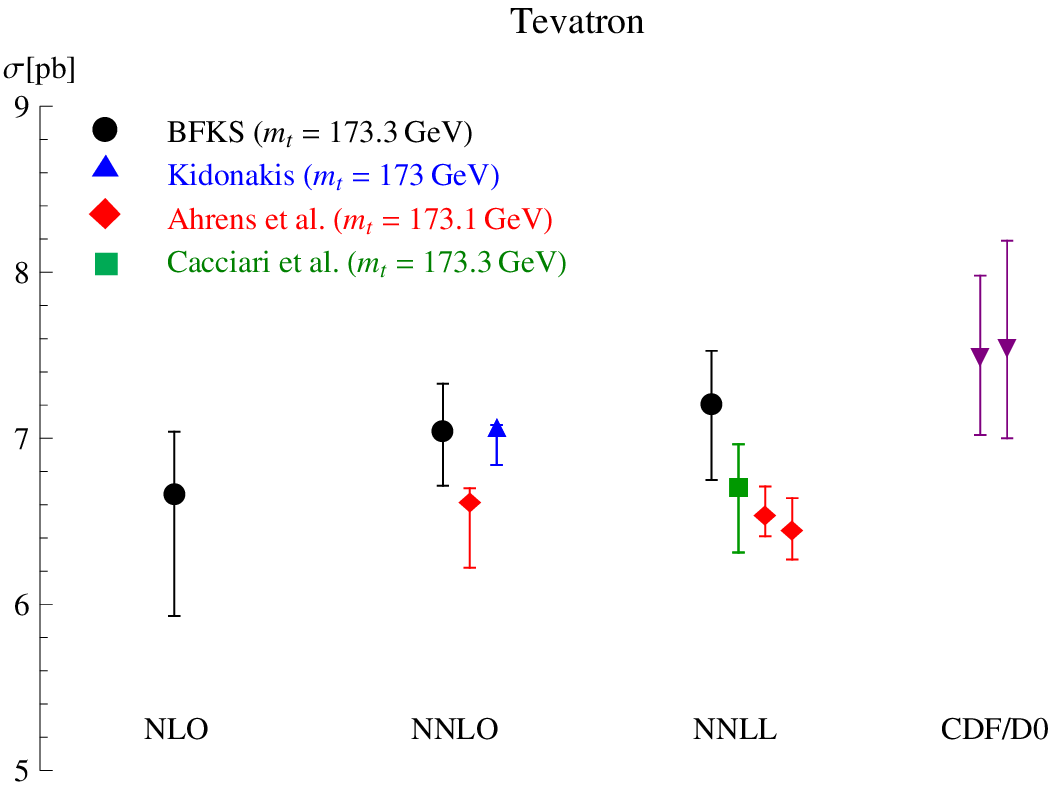}
\includegraphics[width=0.48 \linewidth]{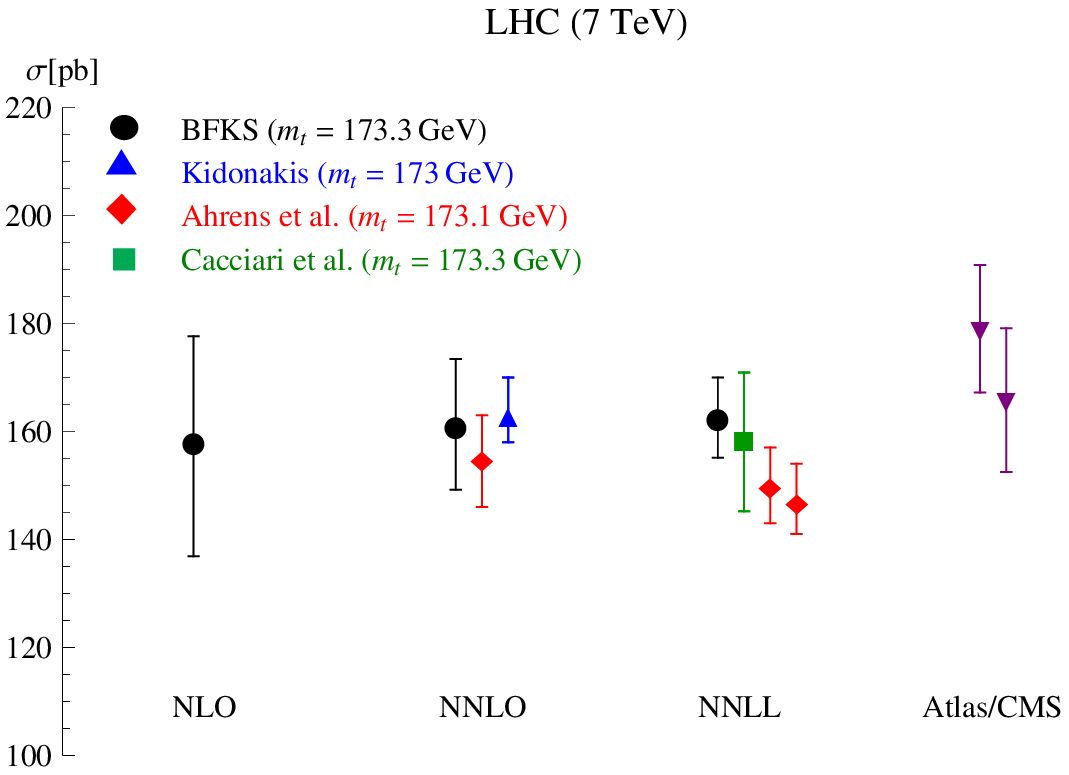}
\end{center}
\caption{Comparison of different NNLO and NNLL predictions, see the text for explanation and references.
The error bands include theoretical uncertainties, but no PDF+$\alpha_s$ errors. The rightmost set of points represents the most recent experimental measurements, which assume a top mass of 172.5 GeV.}
\label{fig:comp}
\end{figure}

Predictions for the total cross section at the NNLL or approximated NNLO level have been published 
recently by several groups. These results differ in the resummation formalism adopted and in whether the total cross section is resummed, 
or differential distributions in different kinematics
limits, 
which are then integrated to obtain predictions for the inclusive cross section.
They also differ in the treatment of Coulomb effects beyond NLO and of the constant terms at ${\cal O}(\alpha_s^4)$, and some include sets of power-suppressed
contributions in $\beta$. 
A comparison of the various results can thus give an estimate of the ambiguities inherent to resummation. 
This is shown in Figure \ref{fig:comp}, where the predictions for NNLO and NNLL given here (black circles, \cite{arXiv:1109.1536})
are compared to the numbers by Kidonakis (blue up-pointing triangles, \cite{arXiv:1109.3231}), Ahrens et al. (red diamonds, \cite{Ahrens:2011px} for NNLO and
\cite{Ahrens:2011mw} for NNLL in both 1PI and PIM kinematics) 
and Cacciari et al. (green squares, \cite{arXiv:1111.5869}). The experimental measurements (purple
down-pointing triangles, \cite{arXiv:1004.3224,arXiv:1101.0124,Aad:2011yb,Chatrchyan:2011yy}) are also
given for comparison. At the LHC the predictions of different groups show a good agreement, with a somewhat 
better agreement for the NNLO results. At NNLL, the total spread of the four results is still smaller than the 
theoretical uncertainty of the NLO result. At Tevatron there appears to be a stronger tension between different results (both for central values and 
error estimates), with the envelope of all predictions at NNLO and NNLL being almost as large as the uncertainty of the NLO result. 
In \cite{arXiv:1109.1536} it was argued that this might be a consequence of the dominance, at Tevatron, of the $q \bar{q}$ production channel, 
which is less well approximated by its threshold expansion than the $gg$ channel dominant at the LHC, thus leading to larger ambiguities in resummation.    

\section{Mass determination}

\begin{figure}[t!]
\begin{center}
\includegraphics[width=0.5 \linewidth]{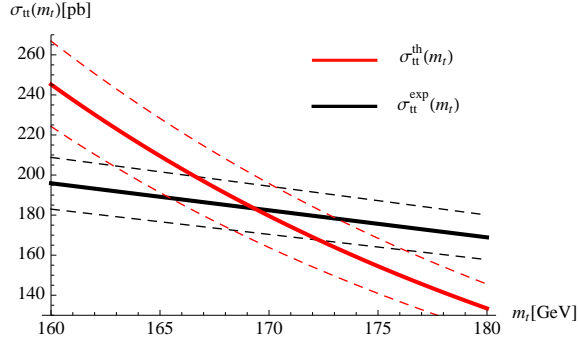}
\end{center}
\caption{$m_t$ dependence of the experimental $t\bar{t}$ cross section \cite{Aad:2011yb} (solid black) and of the NNLL$_2$ 
approximation presented here (solid red). The dashed lines represent the total uncertainties of the two curves.} 
\label{fig:mass}
\end{figure}

Precise theoretical predictions of $\sigma_{t \bar{t}}$ can be translated into measures of the top-quark pole mass $m_t$.
In first approximation, this can be done by comparing the mass dependence of the theoretical and experimental cross sections, 
and extracting the top mass at the intersection point. In a more sophisticated approach, one defines a likelihood function 
\begin{equation} \label{eq:mass}
f(m_t) \propto \int d \sigma \,f_{\text{th}}(\sigma|m_t) \cdot f_{\text{exp}}(\sigma|m_t) \, ,
\end{equation}   
where $f_{\text{th}}$ and  $f_{\text{exp}}$ represent normalized gaussian distributions
centred around the theoretical prediction and measured experimental cross section, respectively,
and extracts $m_t$ from the maximum of the probability distribution. 
A suitable parameterization of the experimental cross section in terms of $m_t$ has been provided 
by the ATLAS collaboration \cite{Aad:2011yb}, and is plotted in Figure \ref{fig:mass}. Using the NNLL$_2$ prediction presented here as
theoretical input in (\ref{eq:mass}) and the data from \cite{Aad:2011yb} gives the following value for the top pole mass,
\begin{equation}\label{eq:mass_ext}
m_t = 169.8^{+4.9}_{-4.7} \, \text{GeV} \, ,
\end{equation}
which agrees with the direct-reconstruction measurement of Tevatron, $m_t=173.3 \pm 1.1 \, \text{GeV}$.
Note that the error in (\ref{eq:mass_ext}) does not include the
uncertainty deriving from identifying the Monte Carlo mass in 
the experimental input with the pole mass $m_t$. Allowing for a $\pm 2\,$GeV difference would correspond to an additional
uncertainty of $\pm 0.65\,$GeV on the extracted mass. 

\section{Conclusions}

We presented new predictions \cite{arXiv:1109.1536} for the $t
\bar{t}$ total production cross section which include NNLL resummation
of soft and Coulomb effects. Resummation increases the cross section relative to the fixed-order NLO result, and leads to a significant reduction of the theoretical 
uncertainty. Our numbers are in good agreement with experimental measurements at the Tevatron and LHC. At the LHC they also show a reasonably
good agreement with other NNLL predictions, though bigger differences are found at the Tevatron.

\end{document}